\documentclass[conference]{IEEEtran}
\IEEEoverridecommandlockouts
\usepackage{cite}
\usepackage{amsmath,amssymb,amsfonts}
\usepackage{graphicx}
\usepackage{textcomp}

\usepackage{booktabs} % For formal tables
\usepackage{arydshln}
\usepackage{subcaption,verbatim,engord,enumitem,multirow,bm,caption}
\usepackage{algorithm,algorithmic}
\usepackage{amscd}
\usepackage{mathtools}
\usepackage{balance}

\usepackage{xcolor}
\def\BibTeX{{\rm B\kern-.05em{\sc i\kern-.025em b}\kern-.08em
    T\kern-.1667em\lower.7ex\hbox{E}\kern-.125emX}}
\begin{document}

% AUTHOR: Enter the title, all letters in upper case
\title{Sensitivity Analysis in the Dupire Local Volatility Model with Tensorflow}

% AUTHOR: Enter the authors of the article, see end of the example document for further examples
\author{Francois Belletti, Davis King, James Lottes, Yi-Fan Chen, John Anderson\\ [12pt]
Google Research\\
Mountain View CA\\
USA\\
belletti@google.com
}

\maketitle

\begin{abstract}
In a recent paper~\cite{belletti2020tensor}, we have demonstrated how the affinity between TPUs and multi-dimensional financial simulation resulted in fast Monte Carlo simulations that could be setup in a few lines of python Tensorflow code. We also presented a major benefit from writing high performance simulations in an automated differentiation language such as Tensorflow: a single line of code enabled us to estimate sensitivities, i.e. the rate of change in price of financial instrument with respect to another input such as the interest rate, the current price of the underlying, or volatility. Such sensitivities --- otherwise known as the famous financial “Greeks” --- are fundamental for risk assessment and risk mitigation.
In the present follow-up short paper, we extend the developments exposed in~\cite{belletti2020tensor} about the use of Tensor Processing Units and Tensorflow for TPUs.
\end{abstract}


\section{Introduction}
Our aim here is to delve more into Tensorflow as a tool for sensitivity analysis. To that end, we reproduce results produced in~\cite{savine2018modern} on the sensitivity analysis of the local volatility model with Automated Adjoint Differentiation (AAD) (known in the ML community as back-propagation).
Using Tensorflow~\cite{abadi2016tensorflow}, (which enables automated differentiation and makes leveraging GPUs and TPUs~\cite{jouppi2017datacenter} extremely simple), we aim to reproduce quantitative results presented in Section 12.4 (page 424) of~\cite{savine2018modern}.
We demonstrate in particular that on TPUs, in spite of the mixed numerical precision, we are able to closely reproduce results obtained on CPU with standard IEEE float32 precision.
First, we recall the context in which Tensorflow can be used for Monte Carlo simulation in quantitative finance.

\section{Simulating to assess risk}
Most financial assets are subject to frequent and unpredictable changes in price. Quantifying the potential outcomes associated with fluctuations in value of the instruments underlying a financial portfolio is of primordial importance to monitor risk exposure. Monte Carlo simulation is routinely used throughout the financial sector to estimate the potential changes in value of financial portfolios over a certain time horizon. In particular, if multiple instruments in a given portfolio have correlated fluctuations and/or if the portfolio comprises derivatives that share common underlyings, Monte Carlo simulation helps unravel the different outcomes that emerge out of such complex correlation and composition structures.
As variance is particularly large for many financial assets, in general, it is necessary to sample many times (hundreds of thousands to billions) to obtain converged estimates under the Law of Large Numbers. The issue is even more pronounced in high dimensional settings where the use of Quasi Random Numbers~\cite{glasserman2013monte,pages2018numerical,savine2018modern} to speed up convergence is less practical. In short, we routinely need to run the same high dimensional simulation of a stochastic process many times.

\begin{figure}[ht]
    \centering
    \includegraphics[width=0.85\linewidth]{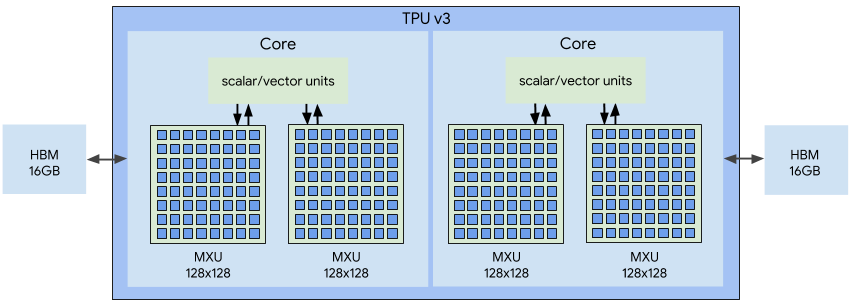}
    \includegraphics[width=0.6\linewidth]{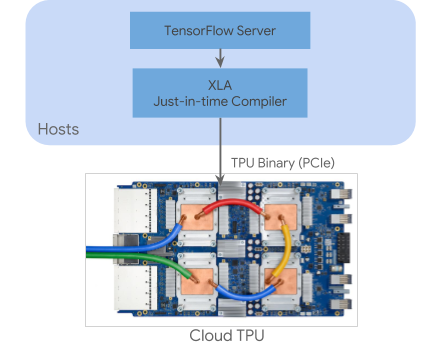}
    \caption{Hardware architecture and programming model of Tensor Processing Units (TPUs)~\cite{cloudtpu}.}
    \label{fig:TPU}
\end{figure}

\begin{figure}
    \centering
    \includegraphics[width=0.95\linewidth]{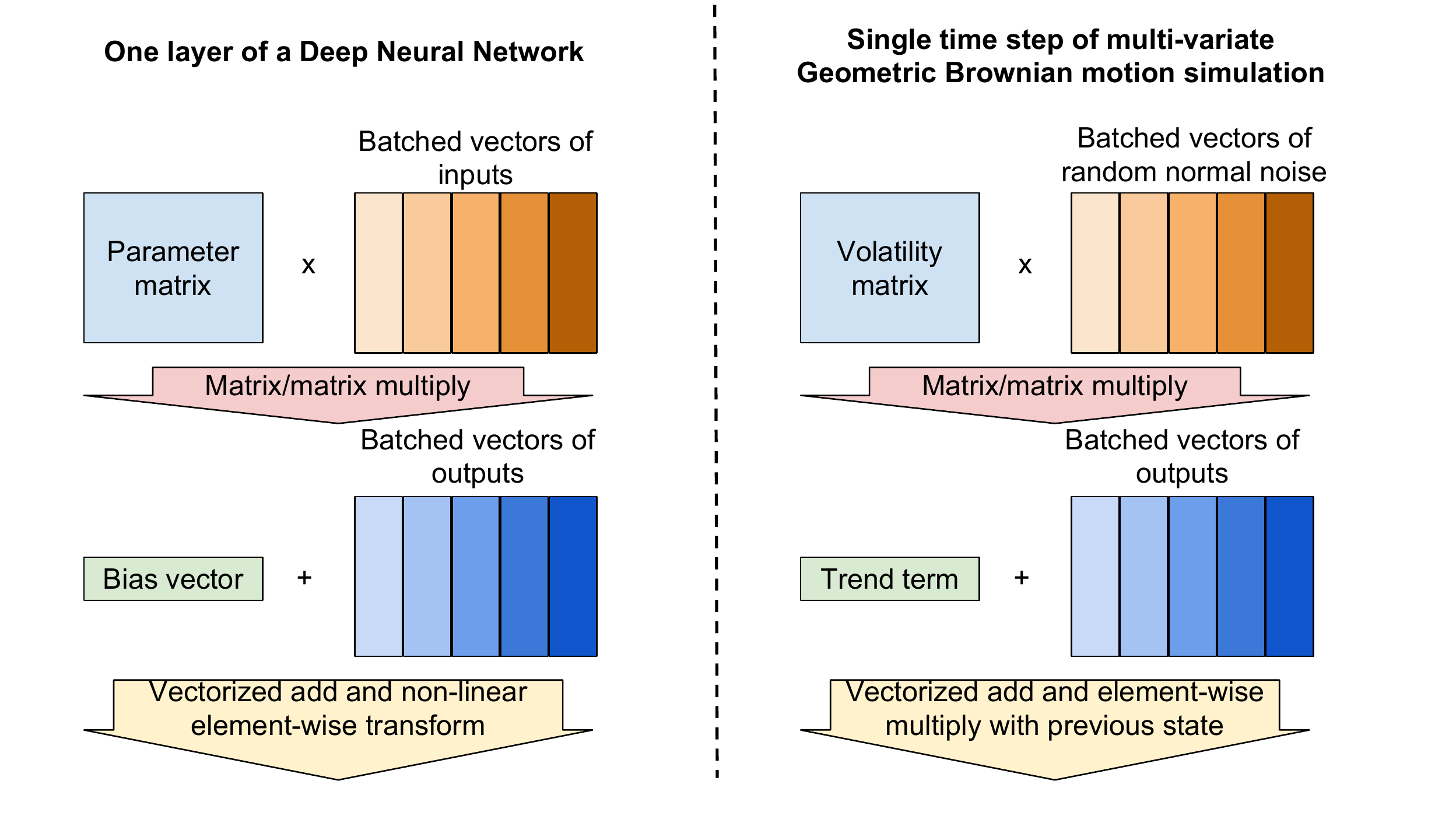}
    \caption{Similarity between a DNN layer and a time step of a multi-dimensional Geometric Brownian Monte Carlo.}
    \label{fig:computational_similarity}
\end{figure}

Running the same program affected by random perturbations over and over again is now familiar to most ML researchers. Stochastic Gradient Descent, which is now a cornerstone of Deep Learning~\cite{goodfellow2016deep}, runs the same computation millions to billions of times with different inputs and subjected to different random perturbations such as dropout or data augmentation techniques.
Recently, Tensor Processing Units~\cite{jouppi2017datacenter,cloudtpu,jouppi2017tpuperf} (see Figure~\ref{fig:TPU}) have been designed to accelerate the training of deep neural networks (in particular dense and convolutional feed-forward networks) which in turns is strikingly similar in terms of computational patterns to multi-dimensional Monte-Carlo simulations that are employed to assess financial risk. Indeed, both rely on interleaving element-wise operators with matrix/matrix products as illustrated in Figure~\ref{fig:computational_similarity}.

In~\cite{belletti2020tensor}, we have demonstrated the efficiency of Tensorflow and TPUs to price European options with a single underlying, price European basket options and compute their ``delta'', estimate Value-at-Risk and Conditional Value-at-Risk, and price American options. In particular, sensitivity analysis was limited to estimating sensitivities with respect to the current price of the underlyings. We now investigate the estimation of sensitivities with respect to model parameters such as the local volatility surface in Dupire's model as in~\cite{savine2018modern}. 

\section{Automated differentiation for financial risk assessment}
In this follow-up paper, we focus closely on the general use of Tensorflow and TPUs to estimate financial Greeks~\cite{hull2003options}. In particular, we show how to estimate the sensitivity of an estimate for the price of an option with respect to a parameter of key importance: the local volatility of its underlying asset.
In other words, we want to understand how robust the price estimate we obtain is with respect to errors in the estimation of the volatility parameter.
Such a procedure is key to understanding the risk associated with model parameter under or over estimation.

A standard technique to estimate such a sensitivity used to rely on the computation of an empirical numerical first order derivative through “bumping”. For each parameter of interest, $p$, a simulation is run with a value of $p_0 + \frac{\epsilon}{2}$, another with $p_0 - \frac{\epsilon}{2}$ and the difference between the two outcomes normalized by $\epsilon$ would serve as an estimate for the sensitivity of the option price with respect to $p$ at the value $p=p_0$. In the case of local volatility model, where the volatility surface comprises hundreds of parameters, the procedure requires twice as many converged simulations as there are parameters to compute sensitivities with respect to.

Recently, the use of Automated Adjoint Differentiation (AAD), i.e. back-propagation, has provided risk assessment with faster means of estimation for sensitivities. Such a first order derivative is practically very easy to implement with the use of an automated differentiator to program the simulation. The advantages of AAD over bumping are exposed at length in~\cite{savine2018modern}.

\section{TPUs to estimate first order derivatives of option price estimates}
Methods other than AAD, such as computing the tangent process of the simulation or employing Malliavin calculus~\cite{pages2018numerical,glasserman2013monte}, can be employed to estimate sensitivities. In all these cases, a simulation of the adjoint, tangent or Malliavin weight is needed which is also largely simplified by automated differentiation as they rely on the calculation of the first order derivative of the transition operation of the discretized stochastic process under study.

All the methods above benefit from two major advantages provided by the use of TPUs.
First, simulations are directly written with an automated differentiation language.
Second, many simulations (and first order derivatives) can be computed fast in an embarrassingly parallel manner on TPUs.

We now demonstrate such advantages in practice as we reproduce an application presented in~\cite{savine2018modern} which focuses on a textbook example representative of simulations that are run pervasively when pricing financial derivatives.
As in this reference, we focus on the estimation of the first order derivative of the option price with respect to each parameter of the volatility surface which is of key importance to understand the risk profile of the financial instrument.

\subsection{The Dupire local volatility model}
The Dupire local volatility model considers a single asset (e.g. a stock price) and assumes (once discretized by a naive Euler explicit scheme) that tomorrow’s price equals today’s price affected by a deterministic trend and a crucially important stochastic Gaussian term whose variance depends on today’s stock price. Such volatility depends more precisely on the level of the stock price and the day under consideration.
The corresponding Stochastic Differential Equation can be written as~\cite{savine2018modern} 
$$
\frac{dX_t}{X_t} = \sigma \left(X_t, t\right) dW_t
$$
where $X_t$ is the price of the underlying asset of interest, $\sigma(\cdot, \cdot)$ is the local volatility function and $dW_t$ the Brownian motion representing the process driving price fluctuations.
The volatility surface is classically calibrated thanks to the Black-Scholes model and a root finding procedure for a grid of values of price and time 
$\left\{ 
    \left(
        i \times \Delta x, j \times \Delta t
    \right) 
\right\}_{i = 0 \dots I - 1, j = 0 \dots J - 1}$. An interpolation procedure is then used to output volatility values for any price and time value. Here we consider this calibration has been conducted and we want to understand the sensitivity of the price of a European call option with respect to the volatility surface parameters.
Our experiments employ the very same parameters as our reference (Section 12.4 page 424).
As we conduct our experiment we want to assess multiple points:
Can we implement the Dupire local volatility model efficiently on TPU?
In spite of the use of mixed precision on TPU, can we obtain price and sensitivity estimates that closely match our reference?
Are TPU based simulations and AAD fast enough when compared to the CPU reference wall time given in our reference and with respect to Tensorflow GPU?

\subsection{Implementing the local volatility model on TPU}
We implemented our simulation naively in an interactive notebook using our TF-Risk library~\cite{tf-risk}.
In the code snipped~\ref{fig:code_snipped} provided in appendix, it is noteworthy that we only use the library for three elements:
\begin{itemize}
    \item European call payoff;
    \item A wrapper around pseudo random normal number generators as provided by Tensorflow;
    \item A 2d interpolation method optimized for TPUs.
\end{itemize}

\subsection{Accuracy of TPU-based estimates}
As a parameter, we employ a volatility surface which --- as in our reference --- comprises 30 discretization points for price values and 60 discretization points for time values.
We want to compute the sensitivity of the estimated price (which in practice is estimated through Monte Carlo sampling and averaging) with respect to each of the 1800 parameter values of the volatility surface.
Tensorflow enables us to write such a procedure in a few lines of code while implicitly conducting automated differentiation through back-propagation to estimate the average first order derivative of the price estimate with respect to each volatility parameter.

\begin{figure}[ht]
    \centering
    \includegraphics[width=\linewidth]{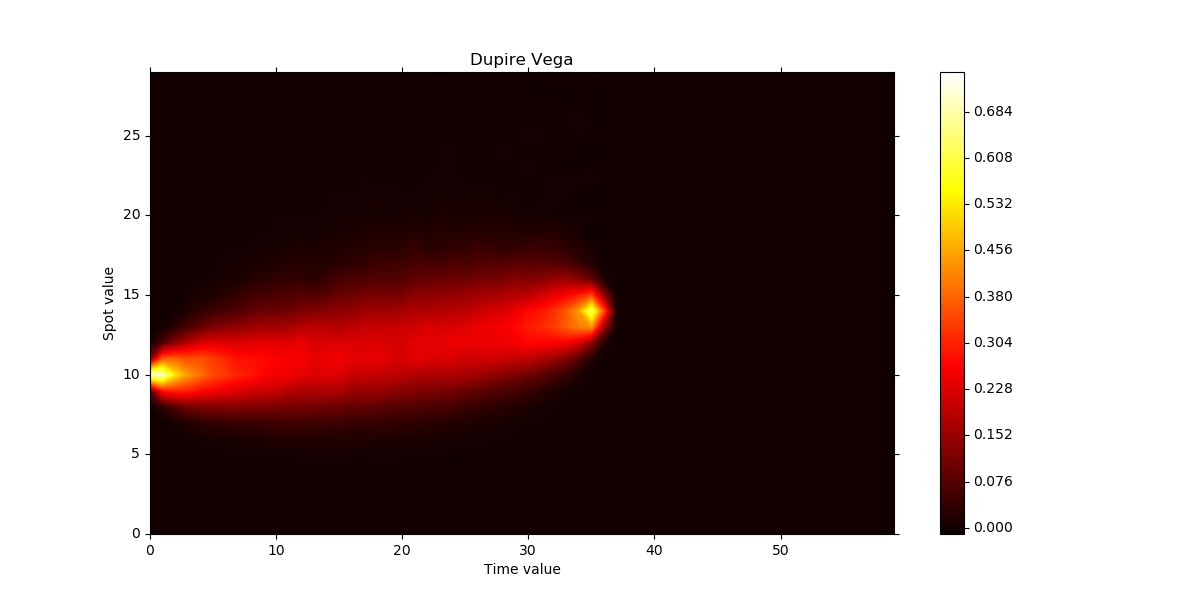}
    \caption{Estimated sensitivities of the European call price with respect to the volatility surface parameters as in~\cite{savine2018modern}.
    \label{fig:sensitivity_surface_smooth}}
\end{figure}

\begin{figure}[ht]
    \centering
    \includegraphics[width=0.9\linewidth]{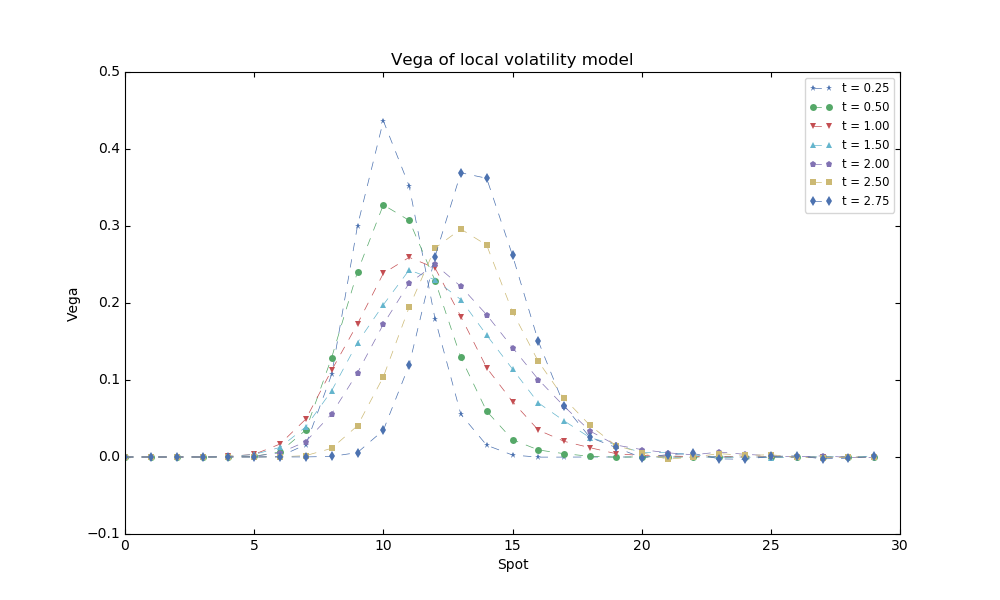}
    \caption{
Estimated sensitivities match the results presented page 425 in~\cite{savine2018modern} closely (our results are not smoothed).
\label{fig:sensitivity_estimates}}
\end{figure}

We can plot each of these 1800 estimated first order derivatives as usually do for the volatility surface. The corresponding sensitivity estimates are presented in Figure~\ref{fig:sensitivity_surface_smooth}. 
Furthermore, we can accurately reproduce the results presented in~\cite{savine2018modern} with 500K sampled trajectories as shown in Figure~\ref{fig:sensitivity_estimates}.

\subsection{TPU based estimation speed}
It is reported in~\cite{savine2018modern} that the high performance parallel C++ library built by the authors produced estimates with $500$K sampled trajectories in $575$ms at best which gives us a state-of-the-art reference for multi-threaded CPU implementation.
After compilation of our python code for TPU with XLA, we report a median wall time of $16$ms from an interactive notebook backed by a full Cloud TPU v3 with computations in mixed precision (float32 for element-wise operations and bfloat16 for operations involving the MXU). Our wall time comprises the round-trip necessary for the notebook’s kernel to send instructions to the TPU and get results back.
We also compare against a Tensorflow GPU implementation and there the best wall time is 110ms on a V100 GPU. 

It is noteworthy that we optimized our implementation of the two dimensional interpolation routine employed in the simulation to compute the value of the local volatility between spots and times for which it has been estimated (our price can take any positive value while we have 156 simulated time steps and only 60 maturities for which the local volatility has been estimated). As in~\cite{savine2018modern}, we employ a bi-linear interpolation which creates a differentiable path between the estimated parameters and the option price estimate.
Bi-linear interpolation can be implemented in various ways and a naive method consists in looking up values through a $tf.gather\_nd$. We found that this implementation was suboptimal both for V100 GPUs and TPUs. Indeed, in both cases, it was preferable to employ one-hot encodings and matrix/matrix multiplies which in turn could help us leverage the TensorCore unit on V100s and the MXU on TPUs.

\section{Conclusion}
Although TPUs have limited numerical precision, they can successfully be leveraged with Tensorflow to conduct reliable general purpose sensitivity analysis for financial derivatives which in turn improves quantitative risk assessment.
One line of code is sufficient, once a simulation has been written in Tensorflow, to compute first order financial ``Greeks''.
Furthermore, Tensorflow readily enables the use of GPUs or TPUs in the cloud without substantial code changes which in turn enables strong wall time improvements for the computation of simulations and sensitivities when compared to CPUs.

Additionally, Tensorflow provides facilities to scale up simulations to multiple CPU/GPU machines or entire TPU-pods in the cloud, which we will show in our upcoming work. We will also demonstrate how to provide support for near double numerical precision on TPUs.

\section*{Appendix A: code snippet}
\begin{figure}[H]
    \centering
    \includegraphics[width=\linewidth]{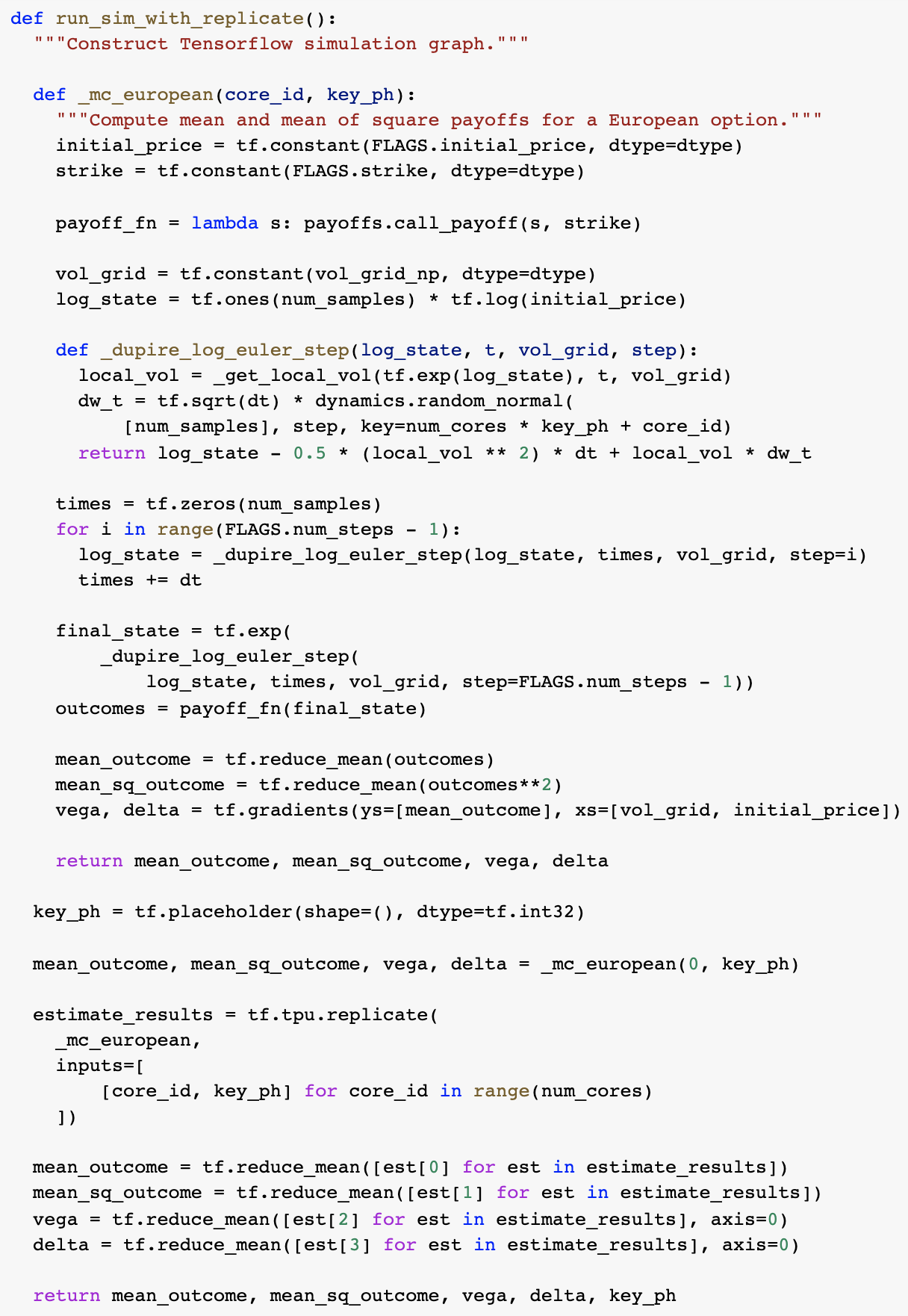}
    \caption{Implementation of Dupire's local volatility model with TF-Risk.}
    \label{fig:code_snipped}
\end{figure}

In Figure~\ref{fig:code_snipped}, we showcase the python Tensorflow implementation of Dupire's local volatility model in TF-Risk~\cite{tf-risk}.

\bibliographystyle{acm}
{\small
\bibliography{main}
}

\end{document}